**Holographic intravital microscopy for 2-D and 3-D imaging intact circulating blood cells in microcapillaries of live mice**

Kyoohyun Kim[1,+], Kibaek Choe[2,+], Inwon Park[3], Pilhan Kim[2,*] and YongKeun Park[1,4,*]

[1] Department of Physics, Korea Advanced Institute of Science and Technology, Daejeon 34141, Republic of Korea

[2] Graduate School of Nanoscience and Technology, Korea Advanced Institute of Science and Technology, Daejeon 34141, Republic of Korea

[3] Graduate School of Medical Science and Engineering, Korea Advanced Institute of Science and Technology, Daejeon 34141, Republic of Korea

[4] TomoCube, Inc., Daejeon 34051, Republic of Korea

[+] These authors contribute equally to this work.

*Correspondence:

Prof. YongKeun Park,

E-mail: yk.park@kaist.ac.kr,

Tel: +82-42-350-2514, Fax: +82-42-350-2510

Prof. Pilhan Kim

E-mail: pilhan.kim@kaist.ac.kr,

Tel: +82-42-350-1115, Fax: +82-42-350-1110

**Intravital microscopy is an essential tool that reveals behaviours of live cells under conditions close to natural physiological states. So far, although various approaches for imaging cells *in vivo* have been proposed, most require the use of labelling and also provide only qualitative imaging information. Holographic imaging approach based on measuring the refractive index distributions of cells, however, circumvent these problems and offer quantitative and label-free imaging capability. Here, we demonstrate *in vivo* two- and three-dimensional holographic imaging of circulating blood cells in intact microcapillaries of live mice. The measured refractive index distributions of blood cells provide morphological and biochemical properties including three-dimensional cell shape, haemoglobin concentration, and haemoglobin contents at the individual cell level. With the present method, alterations in blood flow dynamics in live healthy and sepsis-model mouse were also investigated.**

**Introduction**

Intravital microscopy (IVM) visualizes intact cells and tissues in live animals with high spatial resolution and has been used as an invaluable tool for studying diverse pathophysiology ranging from cell biology[1] and immunology[2, 3] to tumour biology[4, 5]. Various imaging techniques have been developed and used. Confocal scanning microscopy and multi-photon microscopy have been commonly used for imaging living cells and tissues *in vivo*[6]. Even though these microscopic techniques provide the 3-D shapes of cells and their dynamics, they have limited capability to quantifying physiological information about cells. Moreover, the use of exogenous labelling agents significantly limits the use of these techniques only to animal studies. In clinics, bright-field intravital microscopes have been widely used in clinics because they do not require the use of labelling agents. However, they only provide limited imaging capabilities – two-dimensional (2-D) qualitative imaging information.

In this work, we claim that holographic approach can circumvent these limitations in conventional IVM because it can quantitatively measure refractive index (RI), intrinsic optical properties of biological cells and tissues. Recently, quantitative phase imaging (QPI) techniques have emerged to yield quantitative information on individual biological cells and tissues *in vitro*, without using exogenous labelling agents[7]. Exploiting interferometry, QPI techniques measure optical phase delay images of cells, from which various morphological, biochemical, and biomechanical information can be retrieved. Due to its unique capability, QPI techniques have been extensively used for the study of cell pathophysiology, including cell division and growth[8, 9], the morphology of blood cells[10-13], neuroscience[14-16], parasitology[17, 18], and tissue imaging[19, 20].

Due to label-free and quantitative capability, these holographic imaging approaches in QPI have opened the door for several biological and medical studies for cells and tissues *in vitro*. However, it is difficult to perform *in vivo* QPI imaging and for IVM applications, because of light scattering in tissues. Inhomogeneous distributions in tissues cause a significantly large degree of light scattering[21], which scrambles the optical light fields making it difficult to perform holographic imaging *in vivo*. Consequently, QPI for *in vivo* applications has remained unexplored. Despite the many challenges, there is strong motivation to extend QPI to *in vivo* imaging to utilize its unique quantitative and label-free imaging capability, especially if 3-D imaging of individual cells can be measured.

Here, we present intravital QPI imaging of blood cells circulating in the intact microvasculature of a live mouse mesentery. A Mach-Zehnder interferometric microscope measures the complex optical fields of the beam transmitted from the mesentery, from which 2-D quantitative phase maps and 3-D RI tomograms are retrieved. Light scattering effects from adipose tissues surrounding the microcapillaries are filtered out, using the fact that the time-varying signals are from moving blood cells whereas stationary signals are from light scattering by adipocytes. Using the present method, we demonstrate label-free and quantitative 2-D and 3-D imaging of blood cells *in vivo*. Biochemical and morphological properties of individual red blood cells (RBCs) are precisely retrieved, from which fluid dynamics of blood circulation were systematically investigated. Furthermore, we successfully visualized alterations in blood circulation in the microvasculature caused by sepsis.

**Results**

**Optical setup**

In order to perform intravital QPI of microvasculature, we used interferometric microscopy with a method to filter out static scattering signals. A Mach-Zehnder interferometric microscope was used to measure optical fields of a coherent laser beam passing through tissues (Fig. 1a). A diode-pumped solid-state laser ($\lambda$ = 532 nm, 100 mW, Cobolt Co., Sweden) was used as an illumination source. After spatial filtering, the beam from the laser is divided into two arms by a beam splitter. One arm is used as a reference beam. The other beam impinges onto a sample with a long-working distance dry objective lens (LMPLFLN 50×, NA = 0.5, Olympus Inc., Japan) and a tube lens ($f$ = 200 mm). For tomographic measurements, the angle of illumination is rotated by using a dual-axis scanning galvanometer (GVS012, Thorlabs, NJ, USA). Mouse mesentery was staged on a sample plane equipped with a heating plate (Fig. 1b). The diffracted beam from the sample is collected by either a high numerical aperture (NA) objective lens (UPLSAPO, 100×, oil immersion, NA = 1.4, Olympus Inc.) or a water-immersion objective lens (UPLSAPO, 60×, water immersion, NA = 1.2, Olympus Inc.). The beam is further magnified four times with an additional 4-$f$ telescopic system so that total field-of-view becomes 35.0 μm × 35.0 μm (high NA objective lens) or 58.4 μm × 58.4 μm (water-immersion objective lens). The 2-D spatial resolution using the high NA objective lens and the water-immersion objective lens are calculated as 190 nm and 221.6 nm, respectively, which are defined as $\lambda/(2NA)$. The beam diffracted from the sample interferes with the reference beam at an image plane, which generates spatially modulated holograms. The holograms are recorded by a high-speed CMOS camera (1024 PCI, Photron USA Inc., CA, USA) with a frame rate of 1,000 Hz and the exposure time of 1/5,000 sec.

**Field retrieval method from scattering tissues**

From the recorded raw holograms, complex optical field images of a sample, consisting of amplitudes and phase delays maps, were extracted with a field retrieval algorithm[22]. The measured complex optical fields contain signals from scattered light from surrounding tissues as well as light diffracted from blood cells. The scattered light from surrounding adipose tissues is significant because of high RI mismatch between adipocytes and surrounding medium; adipose tissues consist of lipid-storing adipocytes which have high RI ($n \sim 1.46$)[23, 24] compared to that of surrounding medium (for phosphate-buffered saline (PBS) solution, $n = 1.337$ at $\lambda = 532$ nm)[25]. Thus, individual circulating blood cells for our interest are difficult to visualize in the measured optical field images (Fig. 1c), and retrieved phase delays are contributed by both surrounding tissues and circulating cells (Fig. 1d).

To selectively filter out scattered signals from surrounding tissues, we developed a scattering reduction method which is shown in Figs. 1e–f. The principle of the scattering reduction method is the optical field subtraction of stationary signals. The method exploits the fact that even though scattering from surrounding tissue is significant, they are stationary as long as tissues do not move over time because light scattering is a deterministic process[26]. Thus, contributions from light scattering from surrounding tissues can be removed by subtracting individual optical field images by the time-averaged optical field image. Then, the resultant optical fields contain information about time-varying objects, which are circulating blood cells in this case (Fig. 1e). To extract the optical phase delays of RBCs relative to blood plasma, the optical phase delay of blood plasma inside capillaries independent of the flow of RBCs is subtracted, which is obtained as the minimum phase values of sequential frames. The optical phase delay of the RBCs circulating in microvasculature is finally obtained (Fig. 1f). For 3-D tomographic imaging, the scattering reduction method is applied for sequential holograms with the same illumination angle.

**Two-dimensional intravital QPI**

To demonstrate the capability of the presented intravital QPI technique and the scattering reduction method, we first measured 2-D quantitative phase images of live mouse RBCs circulating inside microvasculature in live mouse mesentery, which was prepared in an imaging chamber. Figure 2a and Supplementary Movie 1 show the 2-D optical phase delay of individual RBCs flowing inside a capillary over 0.5 seconds. As individual RBCs are passing through the capillary with a smaller diameter than the size of RBCs, the quantitative phase maps show shape deformation of RBCs inside the capillary. In addition to measuring 2-D optical phase delay of RBCs, the present technique can image other blood components including platelets and white blood cells. For instance, small-sized cells with low optical phase delay indicated in the insets in Fig. 2a are believed to be platelets, which require further investigation. From 2-D optical phase delay of live RBCs, the dry mass of individual cells, a mass of nonaqueous component inside cells, can be calculated[27, 28]. We calculated the dry mass of 106 individual intact RBCs circulating in microcapillaries, and the distribution of measured haemoglobin (Hb) contents of individual RBCs was 16.25 ± 3.06 pg [standard deviation of the mean] (Fig. 2b), which is within the normal range of Hb contents of BALB/c mice[29, 30].

In addition to the quantitative measurement of the biochemical characteristics of individual RBCs from each 2-D phase map, time-lapse phase maps of RBCs flowing inside capillaries provide quantitative information about the fluid dynamics of live RBCs. The velocity of 106 individual intact RBCs circulating inside 7 capillaries was measured from the trajectory of the centre of mass of individual RBCs at each frame. Figure 2c shows that the velocity of RBCs flowing inside microvasculature is within the range of 100 - 700 μm/s, which was reported previously using flow cytometry[31].

The present technique is also capable of measuring quantitative phase delay of a large number of RBCs as well as individual RBCs shown in Fig. 2d and Supplementary Movie 2. By summing the total dry mass of RBCs throughout the field-of-view, the total mass of Hb transported by capillaries can be quantitatively measured, and the temporal change of the total dry mass provides Hb mass flow rates of the bloodstream inside capillaries. The change of flow rates of the bloodstream is further investigated in a bifurcating capillary (Fig. 2e and Supplementary Movie 3), from which the temporal changes of the dry mass of the incoming and outgoing bifurcated capillaries were measured. For incompressible fluid flow, the flow rate of the incoming fluid is the same as the sum of flow rates of the bifurcated outgoing fluid. The flow rate of the bloodstream in bifurcated capillaries, however, shows complicated fluid dynamics because RBCs interact with capillary walls and neighbouring cells, especially in a small capillary (Fig. 2f).

**Quantitative analysis of live blood flow alteration in a sepsis model with intravital QPI**

To extend the applicability of intravital QPI for biological studies, we performed quantitative analysis of the change in blood flow rate in a mouse sepsis model. We intravenously injected lipopolysaccharides (LPS) solution (20 mg/kg), large molecules found in the outer membrane of gram-negative bacteria, to induce a severe septic condition which is a dysregulated systemic inflammatory response including increases in proinflammatory cytokines[32]. We measured time-lapse quantitative phase images before and after the LPS injection for an hour for every 15 minutes, and the same experiment was performed with the injection of PBS solution to compare the effect of the LPS injection on the change in blood flow. As shown in Fig. 3a and Supplementary Movie 4, the elapsed time for individual RBCs to pass through a microcapillary is maintained before (34 ms) and at 45 minutes after PBS injection (30 ms). However, the elapsed time (379 ms) for individual RBCs to passing through a microcapillary was much longer at 45 minutes after the LPS injection compared with that (19 ms) in the same trajectory of the identical capillary before the LPS injection (Fig. 3b and Supplementary Movie 5). The difference in the effect of PBS and LPS injection on blood flow is more clearly seen in the space-time phase images in Figs. 3c-d, where the slope of the phase image of individual RBCs indicates the flow velocity. The flow velocity of RBCs after LPS injection significantly decreases compared to the velocity before the injection, while the change in the flow velocity is minimal in the PBS injection.

For further quantitative analysis, relative volume flow rate and Hb mass rate were measured in 4 and 5 mice for the LPS and PBS groups, respectively (Figs. 3e-f). The relative volume flow rate and Hb mass rate in LPS injected group decreased quickly compared with those in the PBS injected group. These results indicate insufficient blood supply to peripheral tissue in LPS-induced sepsis. Decreased blood flow in capillaries is one of the well-known microcirculatory alterations in sepsis which causes various organ failures by insufficient supply of oxygen. Decreased systemic pressure and local arteriolar constriction in sepsis cause blood flow reduction in microcapillaries[33, 34]. In addition, the blood flow reduction in microcapillaries may also be a result of decreased RBC deformability in sepsis which has been described by several mechanisms: LPS or inflammatory cytokines-induced high production of nitric oxide in sepsis causes decreased RBC deformability through increased intracellular $Ca^{2+}$ concentrations in RBCs; reduction of intracellular ATP in RBCs in sepsis also causes increased intracellular $Ca^{2+}$ concentrations in RBCs through decreased energy for $Ca^{2+}$ membrane pump, and excessive amount of reactive oxygen species (ROS) released by white blood cells can alter RBC membranes[34].

**Three-dimensional intravital QPI**

The expansion of the spatial dimension of QPI from 2-D to 3-D can reveal further quantitative information on biological samples. From 3-D QPI employing optical diffraction tomography (ODT) technique, the present technique measured the 3-D RI distribution of live RBCs in blood flow, which provides quantitative structural and biochemical information on the samples including protein concentration, dry mass, cell volume, surface area, and sphericity[12, 35].

Figure 4 and Supplementary Movies 6-7 show 3-D RI distribution of individual RBCs circulating in a capillary. Cross-sectional slices of 3-D RI tomograms of individual RBCs show that RBCs have parachute shapes in the capillary (Figs. 4a and 4c) while some RBCs also maintain a discoid shape with a dimple (Fig. 4b). In addition to the 3-D *in vivo* visualization of RBCs flowing inside the capillary, the structural and chemical parameters of RBCs such as Hb concentration, Hb contents, and RBC volume were extracted from 3-D RI tomograms of individual RBCs (Fig. 4 right panel). The cell volume, Hb contents, and Hb concentration of 13 intact RBCs were measured as 47.23 ± 6.64 fl, 16.80 ± 2.22 pg, and 35.66 ± 2.38 g/dl, respectively, which are within the range of healthy BALB/c mice[29, 30, 36, 37].

The present technique can also visualize label-free the 3-D structure of microvasculature. The standard deviation map of the time-lapse 3-D RI distribution of RBCs reveals the 3-D morphology of a microcapillary (Fig. 4e and Supplementary Movie 8), from which the cross-sectional area of the microcapillary was calculated as 27.76 μm$^2$. The corresponding capillary diameter is 5.94 μm, slightly smaller than the diameter of RBCs from BALB/c mice (6.6 μm)[38], which implies mechanical deformation of RBCs passing through the microcapillary. To the best of our knowledge, this is the first time measuring the quantitative morphological and biochemical parameters of intact individual cells and microcapillaries *in vivo*.

The reconstructed tomograms of RBCs and microcapillary suffer axial shape elongation as a result of the finite scanning angle of the long-working distance of the objective lens (NA = 0.5). While the spatial resolution of the present method is enough to distinguish detailed features of live microvasculature with sizes ranging in the micrometres, it can be further enhanced by using high-NA objective lenses or implementing a complex deconvolution method[39].

**Discussion**

In this paper, we present intravital QPI of blood cells inside the microvasculature of live mouse mesentery. The 2-D intravital QPI clearly shows that the presented technique can measure Hb contents of individual RBCs *in vivo* without exposing them *in vitro*. The present technique also measures complex optical fields of other blood components which are believed to be platelets. Because RI increment values for most proteins are similar at 0.190 ml/g[40], we expect that the present technique can also measure other cells in live tissues including white blood cells[41], platelets, and circulating tumour cells.

The 3-D intravital QPI technique reconstructed 3-D RI tomograms of capillaries of live mouse mesentery, from which structural and chemical parameters of individual RBCs were extracted. Because the RI of biological samples is an intrinsic optical parameter which is sensitive to morphological structures and chemical compositions, the present technique has a possibility for diagnosing live biological tissues and cells under various pathophysiological conditions. While RI distribution itself does not give molecular specificity for identifying specific proteins in biological samples, measuring RI dispersion of biological cells by spectroscopic phase microscopy can distinguish 3-D distribution of various types of proteins label-free[42-44].

The time-lapse optical phase maps provide quantitative information on fluid dynamics of RBCs as well as Hb mass flow rates inside microvasculature, which is an important quantitative measure for oxygen delivery in the circulatory system. From the time-lapse 2-D optical phase maps, the effect of LPS on blood flow rate was quantitatively investigated by analysing the volumetric flow rate and Hb mass rate of microvasculature in live mice. While conventional techniques have measured volumetric flow rate of microvasculature *in vivo* by multiplying the ensemble velocity of a group of RBCs and estimated the cross-sectional diameter of blood vessels[45, 46], it has been difficult to measure the amount of Hb transported through capillaries. Because the ability for oxygen transport of the circulatory system depends on the Hb mass rate related to the number of individual RBCs and intracellular Hb concentration of RBCs, we believe that the present technique opens a new window for quantitative investigations on fluid dynamics and oxygen transport rates of blood flow in live animals in various pathophysiological conditions including RBCs in diabetes[47], sickle cell vasoocclusion[48] and thrombosis[49]. We expect that the presented technique can be used to investigate various disease-related pathological changes in blood cells inside microvasculature *in vivo*.

The present technique successfully filters out light scattering effects in retrieved complex optical fields which are originated from adipocyte tissues in live mouse mesentery. Light scattering from adipocyte tissues is significant. For instance, the reduced scattering coefficient of rat subcutaneous adipose tissue was measured as $\mu_s' = 14.3$ cm$^{-1}$ in the previous research[50] and the mean free path, $l_s$, of a photon in the tissue is 69.9 μm which is calculated as $l_s = (1-g)/\mu_s'$, where anisotropy, $g$, is typically 0.9 in biological tissues. It implies that the mesenteric adipose tissue in the present research suffers at least 2−3 light scattering events. For thicker tissues in which multiple light scattering events can occur[26], we expect that the transmission matrixes (TMs)[51-53] or optical phase conjugation approaches[54-56] need to be utilized. This may further expand the applications of QPI to intravital imaging fields including tumour biology, neurovascular, and cardiovascular biology.

**Methods**

**Animal preparation**

BALB/c mice aged 13 and 20 weeks were used for 2-D and 3-D imaging respectively. Mice were anaesthetized by intramuscular injection of a mixture of Zoletil® (10 mg/kg) and xylazine (11 mg/kg). The level of anaesthesia was continuously monitored during the experiments with a toe pinch and maintained by intramuscular injection of half the dose of initially injected Zoletil-xylazine mixture whenever a response was observed. About 8 cm of the ileum was exteriorized and placed in an imaging chamber. The intestinal loop was surrounded by wet gauze, and the mesentery region was placed on the glass of the chamber. The exteriorized intestine was maintained at 37℃ with a heating pad and a temperature sensor. To keep the exteriorized intestine moist, warm saline was supplied into the gauze in the chamber every 20 min. Because the working distance of the objective lens is limited, mouse mesentery was fixed on a coverslip by adding the PBS solution. Microcapillaries were embedded in adipose tissues stacked in approximately 2 to 3 layers with the thickness of approximately 100 μm. For the mouse sepsis model, lipopolysaccharides (LPS, 20 mg/kg, *E. coli* 055:B5, L2880, Sigma) dissolved in 100 μl PBS was injected into the tail vein of the anaesthetized mice. Only PBS (100 μl) was injected into the control mice. The LPS solution and PBS were prewarmed to about 36℃ immediately before injection. Animal care and experimental procedures were performed under approval from the Animal Care Committee of KAIST (KA2014-01 and KA2015-03). All the experiments in this study were carried out in accordance with the approved guidelines.

**Dry mass calculation**

The dry mass of individual RBCs (i.e., haemoglobin (Hb) content) was calculated from the retrieved optical phase delay as follows[28]:

$$\sigma(x,y) = \frac{\lambda}{2\pi\alpha}\varphi(x,y) \quad (1)$$

$$Hb = \iint \sigma(x,y)dxdy = \frac{\lambda}{2\pi\alpha}\iint \varphi(x,y)dxdy \quad (2)$$

, where $\sigma(x,y)$ is the dry mass density (pg/μm$^2$); $\lambda$ is the wavelength of the illumination beam; $\varphi(x,y)$ is the measured phase map; $Hb$ is the Hb content of individual cells, and $\alpha$ is a RI increment ($\alpha$ = 0.150 ml/g for both oxygenated and deoxygenated Hb at $\lambda$ = 532 nm[25]).

**Flow rate analysis**

The volumetric flow rate and Hb mass rate were obtained from time-lapse phase images of microcapillaries taken for 500 ms with the frame rate of 1,000 Hz. The volumetric flow rate was calculated by measuring the velocity distribution of microcapillaries in the space-time phase image of the microcapillaries. In the space-time phase image along the centre line of a microcapillary, individual RBCs traveling inside the microcapillary are represented as streaks with slopes corresponding to the velocity of the blood flow[57, 58]. The averaged flow velocity, $v$, was determined from the Radon transform of the space-time phase map as follows:

$$v = \frac{\Delta x}{\Delta t} \tan \theta_{max}, \tag{3}$$

where $\Delta x$ and $\Delta t$ is the spatial and temporal resolution of the space-time phase map, respectively, and $\theta_{max}$ is the angle of the RBC streaks where the maximum variance of the Radon transform of the space-time phase map is found. The volumetric flow rate, $Q$, was calculated by multiplying the averaged flow velocity and the cross-sectional area of the capillary by assuming a cylindrical geometry for the microcapillary as follows:

$$Q = vA = v\left(\pi r^2\right) = \pi v \left(\frac{A_{proj}}{2l}\right)^2, \tag{4}$$

where $A$ and $r$ is the cross-sectional area and the radius of the capillary; $A_{proj}$ is the projection area of the capillary at the normal illumination, and $l$ is the length of the centre line of the microcapillary. The Hb mass rate was calculated by the Hb dry mass in the area between the initial (e.g., $p_1$ and $q_1$) and final position (e.g., $p_2$ and $q_2$ in Figs. 3a and 3c, respectively) of the centre line of the microcapillary divided by the elapsed time for an RBC to pass through the centre line. The calculation was repeated for individual RBCs passing through the centre line as many times as possible to avoid cell-to-cell variation.

**Tomogram reconstruction**

3-D RI distribution of samples was reconstructed from optical fields with various incident angles by employing ODT. A dual-axis galvanomirror circularly scanned 10 incident beams with various azimuthal angles with a scanning rate of 10 ms/cycle. The 2-D Fourier spectra of the measured optical fields with various incident angles were mapped into the 3-D Fourier space according to Fourier diffraction theorem[35, 59, 60]. Due to sparse illumination and limited acceptance angles originating from the finite numerical aperture of the objective lens, the resultant 3-D Fourier space has missing information. The missing information was filled by applying an iterative non-negativity constraint algorithm. The 3-D RI distribution of samples was then obtained by applying 3-D inverse Fourier transform. The detailed information about the tomogram reconstruction algorithm via ODT is presented elsewhere[35]. Optical field retrieval and tomogram reconstruction were performed by customized codes in MATLAB® R2014b (MathWorks Inc., MA, USA). RI isosurfaces of individual RBCs and microcapillaries were rendered by commercial software (Tomocube Inc., Korea).

## Acknowledgements

This work was supported by KAIST, KAIST Institute for Health Science and Technology, National Research Foundation (NRF) of Korea (2015R1A3A2066550, 2014K1A3A1A09063027, 2012-M3C1A1-048860, 2014M3C1A3052537, 2012M3A6A4054261), Korea Health Industry Development Institute (HI15C0399) and the Innopolis foundation (A2015DD126). K. K. acknowledges support from Global Ph.D. fellowship from NRF.

## Author Contributions

K. K. and K. C. performed experiments and analysed the data. I. P. developed the mouse sepsis model. Y. P. and P. K. conceived and supervised the project. All the authors wrote the manuscript.

## Competing Financial Interests

The authors declare no competing financial interests

**Figures with Figure Legends**

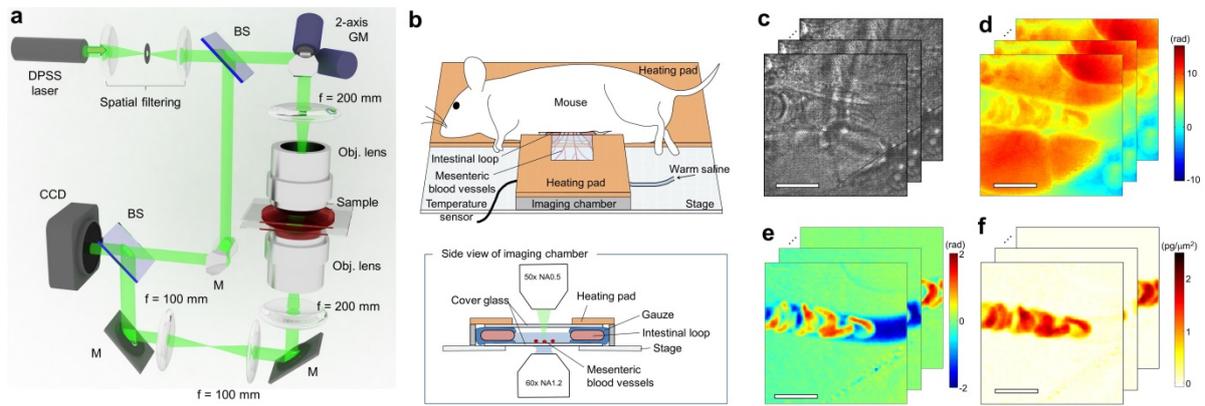

**Figure 1 | Schematic diagram of intravital quantitative phase microscopy. a**, Mach-Zehnder interferometry recording holograms from various illumination angles scanned by a galvanomirror (GM). BS: beam splitter, and M: mirror. **b**, The schematic diagram of live mouse mesentery staged on a heating plate. **c**, Raw holograms recorded by a camera. **d**, Retrieved optical phase delay maps from measured holograms in **c**. **e**, Optical phase delay maps subtracted by the first frame to eliminate surrounding adipocytes. **f**, Final optical phase delay maps subtracted by the minimum values of sequential phase maps which correspond to the phase delay of red blood cells in the microvasculature of the mouse mesentery. Scale bar corresponds to 10 μm.

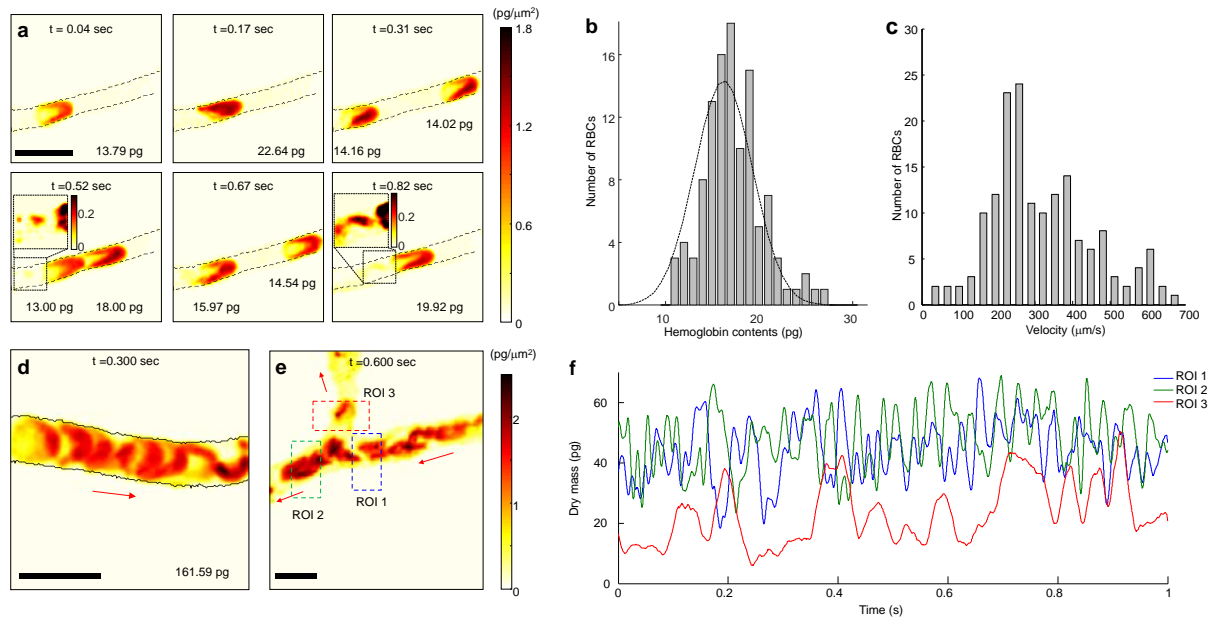

**Figure 2 | Two-dimensional intravital quantitative phase imaging of individual RBCs in microcapillaries. a**, Sequential quantitative phase images of individual RBCs flowing through the microvasculature of the mouse mesentery. Dry mass of individual RBCs is measured from each phase image. Insets are enlarged view of quantitative phase images indicated by dashed boxes, showing blood components other than RBCs. Colorbar indicates dry mass densities, and scale bar corresponds to 10 μm. See also Supplementary Movie 1. **b**, Histogram of the distribution of measured haemoglobin contents of individual RBCs in microcapillaries measured from time-lapse phase maps in **a**. The mean value of the RBCs haemoglobin contents is 16.25 ± 3.06 pg. **c**, Histogram of the velocity distribution of individual RBCs in microcapillaries measured from time-lapse phase maps in **a**. **d**, A snapshot of time-lapse quantitative phase images of the bloodstream flowing inside a capillary. Total dry mass of the field of view (FOV) is measured from the phase image. See also Supplementary Movie 2. **e**, A snapshot of time-lapse quantitative phase images of the bloodstream flowing inside bifurcating capillaries. Colorbar indicates dry mass densities, and scale bar corresponds to 10 μm. See also Supplementary Movie 3. **f**, The change in total dry mass in the regions of interest (ROI) 1, 2 and 3 indicated as boxes in **e**.

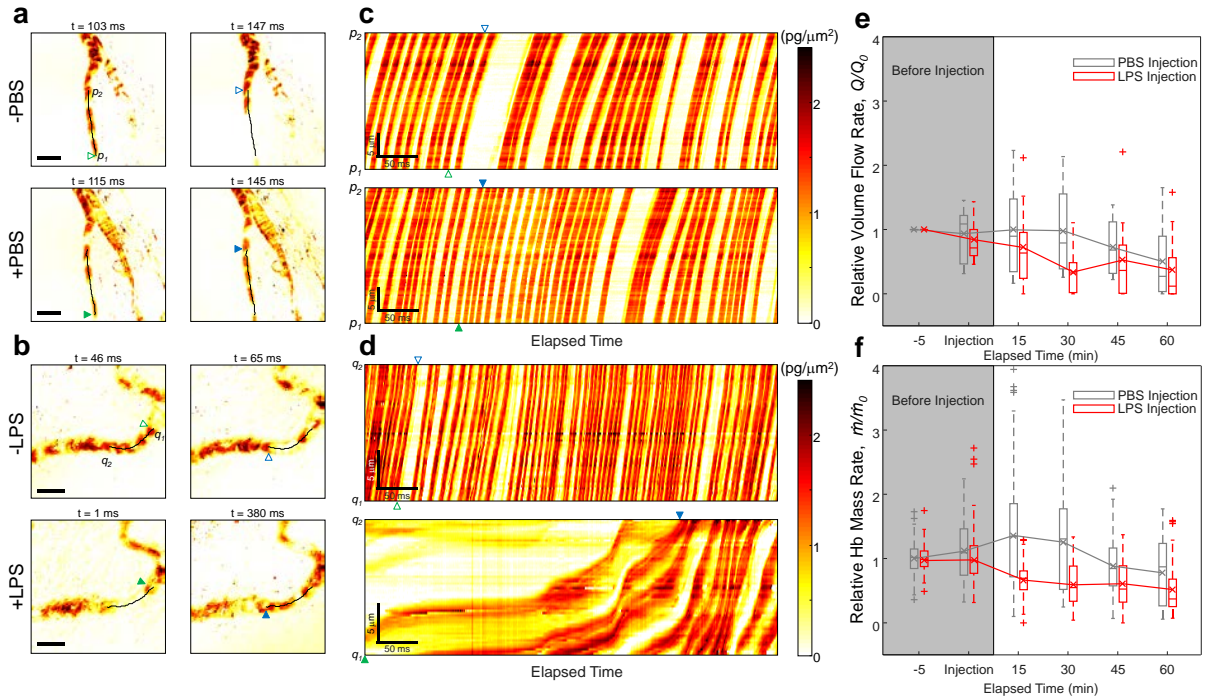

**Figure 3 | Quantitative analysis of the effect of LPS injection on live microvasculature. a-b**, Quantitative phase images of individual RBCs flowing through the microvasculature of the mouse mesentery before (top) and after (bottom panel) the injection of phosphate-buffered saline (PBS) solution (**a**) and lipopolysaccharide (LPS) solution (**b**). Green and blue triangles indicate the initial and final position of RBCs, respectively, along the microcapillaries for which center lines are indicated by solid black lines. Scale bars indicate 10 μm. See also Supplementary Movies 4 and 5. **c-d**, Space-time phase images of RBCs before (top) and after (bottom panel) the injection of PBS solution (**c**) and LPS solution (**d**), respectively. Individual RBCs indicated as green and blue triangles in the phase images in **a-b** are marked as the same triangles. **e-f**, Changes in relative volume flow rate (**e**) and haemoglobin (Hb) mass rate (**f**) over time before and after the injection of PBS solution ($n = 5$, grey boxes) and LPS solution ($n = 4$, red boxes). Values from individual microcapillaries are normalized by the value measured before the injection of the solution.

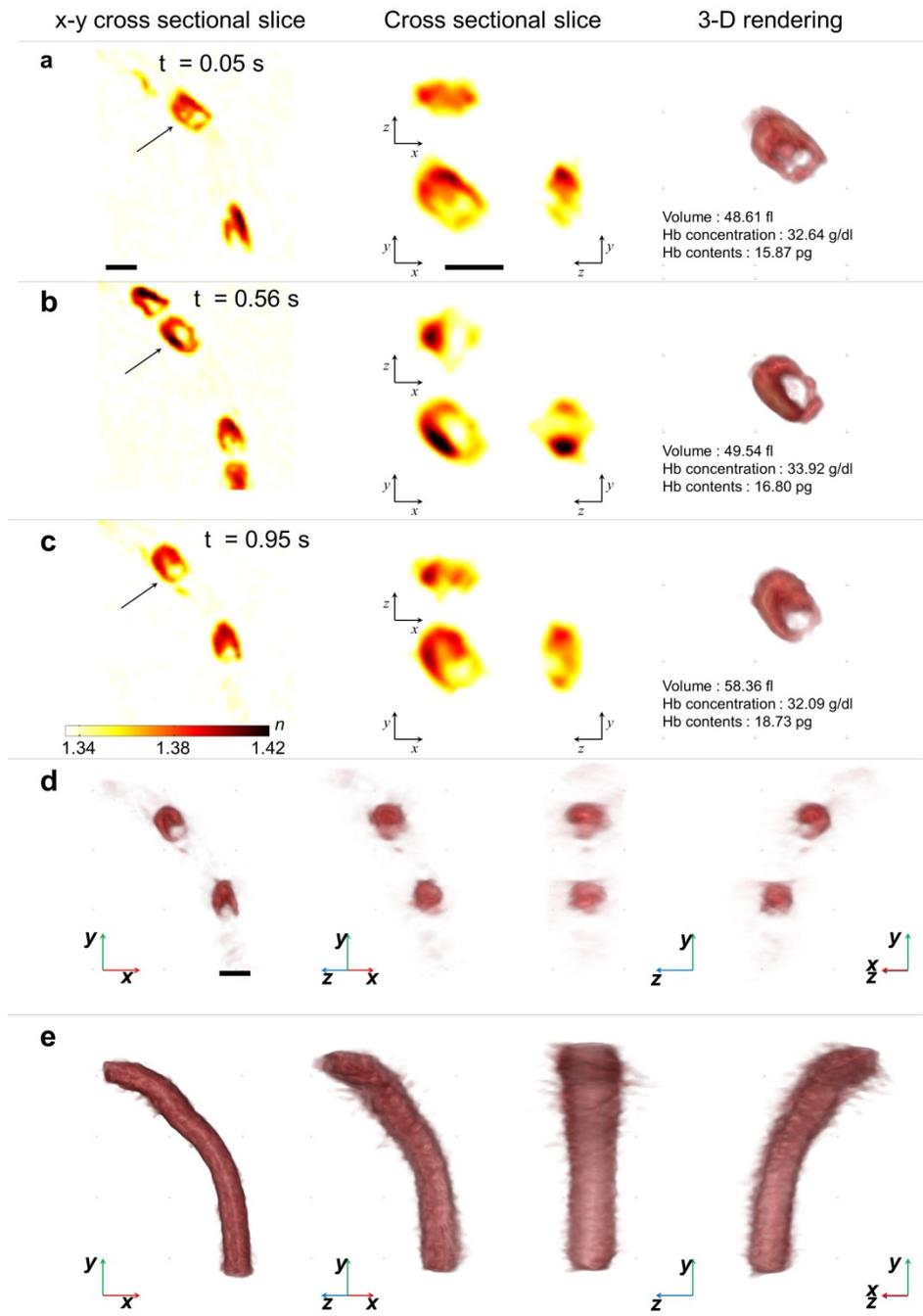

**Figure 4 | Three-dimensional intravital tomographic imaging of individual RBCs in the microvasculature of the mouse mesentery. a-c**, (left panel) *x-y* cross-sectional slices of the three-dimensional RI distribution of individual RBCs flowing through a capillary. See also Supplementary Movie 6. (centre panels) *x-y*, *y-z* and *x-z* cross-sectional slice images of the RI distribution of RBCs indicated by black arrows in the left panel. (right panel) Rendered isosurfaces of the 3-D RI distributions of RBCs and measured volume, haemoglobin concentration and haemoglobin contents. **d**,

Rendered isosurfaces of the 3-D RI distributions of RBCs in the microcapillary in **c** with various viewing angles. See also Supplementary Movie 7. **e**, Rendered isosurfaces of the microcapillary with various viewing angles. See also Supplementary Movie 8. Scale bars indicate 5 μm.